\documentclass[prd,twocolumn,amsmath,amssymb,floatfix,superscriptaddress,nofootinbib,preprintnumbers]{revtex4-1}
 \usepackage[utf8]{inputenc}
\usepackage{graphicx}
\usepackage{amssymb}
\usepackage{amsmath}
\usepackage{bm}
\usepackage{color}
\usepackage{enumitem}
\usepackage{array}

\DeclareMathAlphabet\mathbfcal{OMS}{cmsy}{b}{n}
\definecolor{darkgreen}{RGB}{50,150,0}
\definecolor{purple}{cmyk}{0.5,0.75,0,0}
\definecolor{darkpurple}{RGB}{128,0,128}

\newcommand{\curv}{{\cal R}}
\newcommand{\Planck}{{\it Planck }}

\newcommand{\Comment}[1]{{}}

\definecolor{ultramarine}{rgb}{0.07, 0.04, 0.56}
\definecolor{cadmiumgreen}{rgb}{0.0, 0.42, 0.24}
\definecolor{indigo(dye)}{rgb}{0.0, 0.25, 0.42}
\usepackage[linktocpage=true]{hyperref}
\newcolumntype{L}[1]{>{\raggedright\let\newline\\\arraybackslash\hspace{0pt}}m{#1}}
\newcolumntype{C}[1]{>{\centering\let\newline\\\arraybackslash\hspace{0pt}}m{#1}}
\newcolumntype{R}[1]{>{\raggedleft\let\newline\\\arraybackslash\hspace{0pt}}m{#1}}
\hypersetup{
colorlinks=true,
citecolor=ultramarine,
linkcolor=cadmiumgreen,
urlcolor=indigo(dye),
pdfauthor={},
pdftitle={},
pdfsubject={}
}

\begin{document}
\preprint{YITP-SB-16-18}

\title{Inflationary vs.~Reionization Features from \Planck 2015 Data}

\author{Georges Obied}

\affiliation{Harvard University, Department of Physics, \\Cambridge, MA 02138, USA}

\author{Cora Dvorkin}

\affiliation{Harvard University, Department of Physics, \\Cambridge, MA 02138, USA}

\author{Chen Heinrich}

\affiliation{Jet Propulsion Laboratory, California Institute of Technology, Pasadena California 91109}

\author{Wayne Hu}

\affiliation{Kavli Institute for Cosmological Physics, Department of Astronomy \& Astrophysics,  Enrico Fermi Institute, University of Chicago, Chicago, IL 60637}

\author{V. Miranda}

\affiliation{Center for Particle Cosmology, Department of Physics and Astronomy,\\University of Pennsylvania, Philadelphia, Pennsylvania 19104, USA}

\begin{abstract}
Features during inflation and reionization leave corresponding features in the temperature
and polarization power spectra that could potentially explain anomalies in the Planck 2015 data but require a joint analysis to disentangle.  We study the interplay between these two effects using a model-independent parametrization of the inflationary power spectrum and the ionization history.   Preference for a sharp suppression of large scale power is driven by
a feature in the temperature power spectrum at multipoles $\ell \sim 20$, whereas preference
for a component of high redshift ionization is driven by a sharp excess of polarization power
at $\ell \sim 10$ when compared with the lowest multipoles.   Marginalizing inflationary
freedom does not weaken the preference for $z \gtrsim 10$ ionization, whereas marginalizing
reionization freedom slightly enhances the preference for an inflationary feature but can also mask its direct signature in polarization.
The inflation and reionization interpretation of these features makes predictions for the
polarization spectrum which can be tested in future precision measurements especially at $10\lesssim \ell \lesssim 40$.
\end{abstract}

\maketitle

\section{Introduction}
\label{sec:intro}

Measurements of the anisotropies of the cosmic microwave background (CMB) are entering the era wherein both the temperature and  polarization spectra will be determined to near cosmic variance precision across the linear regime.   Indeed CMB measurements have already helped establish the cosmological constant cold dark matter ($\Lambda$CDM) model with nearly scale invariant inflationary initial conditions  as the standard cosmological model.  Its concordance with other  high precision cosmological probes such as Type IA supernovae, baryon acoustic oscillations and Big Bang Nucleosynthesis  lends strong support for the basic framework of $\Lambda$CDM.

Nonetheless, there are a few well-known tensions and glitches in the CMB power spectra that seem to hint at deviations from the standard cosmological model. With improvement in CMB polarization data in particular, we can search for matching features and consistency tests for potential
physical explanations of these features.

In this work we focus on the large angle features in temperature and polarization power
spectra and test their potential explanation from corresponding features from inflation
and reionization.
The standard $\Lambda$CDM cosmology assumes that the reionization of hydrogen atoms in our universe happened suddenly in redshift, but reionization cannot occur instantaneously and with multiple sources  of ionizing radiation could in principle be quite extended in redshift~\cite{Barkana:2000fd,Ahn:2012sb,Miranda:2016trf}.
Features in the \Planck 2015 polarization data indeed allow and even  favor
a component of ionization at $z\gtrsim 10$ when the complete information out to a limiting redshift is included in a reionization model independent approach \cite{Heinrich:2016ojb,Heinrich:2018btc} (see also \cite{Hazra:2017gtx}).

Incorrect assumptions about the ionization history can bias not only the Thomson total optical depth~\cite{Kaplinghat:2002vt,Holder:2003eb,Colombo:2004uh} but also inferences on parameters not directly related to the physics of reionization such as the sum of the neutrino masses \cite{Smith:2006nk,Allison:2015qca} and tests of the inflationary consistency relation
\cite{Mortonson:2007tb}.  Optical depth constraints when combined with measurements of the heights of the acoustic peaks also impact the gravitational lensing
interpretation of oscillatory residuals in the \Planck temperature power spectrum from $\Lambda$CDM
at high multipole  \cite{aghanim2016,Obied:2017tpd}.

Moreover, if  reionization indeed is more complicated than a steplike transition it can
 in principle impact the
 interpretation of the $\ell \sim 20-40$ features in the temperature power spectrum as well as its confirmation in polarization spectra if the range of features overlap
~\cite{Mortonson:2009qv}.   The interpretation of these features affects the calibration of
the physical size of the sound horizon and hence the inference of the Hubble constant
from the acoustic peaks \cite{aghanim2016,Obied:2017tpd} which is in  tension with the local distance ladder measurements (e.g.~\cite{2018arXiv180101120R}).

These features could potentially indicate transient  violations of slow-roll conditions during the inflationary epoch~\cite{Starobinsky:1992ts,Adams:2001vc,Joy:2007na,Ashoorioon:2006wc,Ashoorioon:2008qr}. Most attempts to model such behavior have been constructed {\it a posteriori} which
makes a statistical interpretation of the significance of deviations from $\Lambda$CDM
difficult to interpret.    Model-independent approaches typically find lower significance per parameter~\cite{Hannestad:2003zs,Shafieloo:2003gf,0004-637X-599-1-1,Shafieloo:2006hs,nicholson2009reconstruction,Dvorkin:2010dn,Dvorkin:2011ui,miranda2015} making the search for matching features in polarization and their disentanglement from reionization
even more important \cite{Mortonson:2009qv,miranda2015,Hazra:2017joc}.

It is also natural to ask whether these or other known features in the temperature power spectrum influence the reionization interpretation of polarization measurements.
For example a suppression of large scale power in the inflationary spectrum would require
a higher optical depth during reionization to produce the same polarization spectrum.
Conversely, the measurement of finite polarization power at the lowest multipoles
requires the existence of horizon scale inflationary fluctuations even beyond the $\Lambda$CDM model \cite{Mortonson:2009xk}.

In this work we combine the model independent
approaches to reionization \cite{Hu:2003gh,Heinrich:2016ojb} and inflation \cite{Obied:2017tpd} to address these issues. We can then draw conclusions about inflationary features while marginalizing the impact of ionization history assumptions and vice versa.  We also examine
the predictions each make for future polarization measurements which could confirm their physical interpretation.

This paper is organized as follows:  we describe the data and models considered in this work
in Sec.~\ref{sec:dataandmodels}, analyze the implications of temperature and polarization features for inflation and reionization in Sec.~\ref{sec:rei_infl}, and discuss
our results in Sec.~\ref{sec:conclusions}.  In Appendix~\ref{sec:appendix}, we illustrate the ability of future cosmic variance limited polarization measurements to improve reionization constraints.

\section{Data and Models}
\label{sec:dataandmodels}

In the analysis presented in this work, we use the {\it Planck} 2015 CMB power spectra as provided in the publicly available likelihood functions. Since reionization constraints and their impact on the large scale curvature power spectrum rely mainly on polarization spectra, we use the low multipole ($2\le \ell\le 29$) \texttt{lowTEB} and the high multipole ($\ell \ge 30$) joint TTTEEE \texttt{plik} likelihood.  Parameter inference in each of the models is carried out with a modified version of the Boltzmann solver \texttt{CAMB}~\cite{Lewis:1999bs,Howlett:2012mh} linked to the Markov Chain Monte Carlo public code \texttt{COSMOMC}~\cite{Lewis2002,Lewis2013}.

The baseline case is the standard $\Lambda$CDM model with parameters
$\Omega_b h^2$ for the baryon density, $\Omega_c h^2$ for the cold dark matter density,
$\theta_{\rm MC}$ for the effective angular scale of the sound horizon, and $\tau$ for the total Thomson optical depth. In this baseline model, the inflationary initial conditions are parametrized by
the amplitude $A_s$ and tilt $n_s$ of the curvature power spectrum $\Delta_{\cal R}^2 = A_s (k/k_0)^{n_s-1}$ where the pivot scale is $k_0= 0.08$ Mpc$^{-1}$.
Hydrogen reionization is taken to be a steplike transition and parametrized by the total optical
depth $\tau$ through reionization (see Ref. \cite{Heinrich:2016ojb} for details).

We investigate models which provide more freedom in the reionization and inflationary
histories. For the former, we focus on the average ionization fraction $x_e(z)$, as appropriate for  large angle  power spectra.
Here we allow arbitrary variations around a fiducial model of the ionization fraction \cite{Hu:2003gh}:
\begin{align}
  x_e(z) = x_e^{\rm fid}(z) + \sum_a t_a S_a(z),
  \label{eqn:reionPC}
\end{align}
where $S_a(z)$ are the principal components of the cosmic variance limited $EE$ Fisher matrix for $x_e$ perturbations between $z_{\rm min}=6$ and  $z_{\rm max}$ as constructed in Ref.~\cite{Mortonson:2007hq}, $t_a$ are their amplitudes and $x_e^{\rm fid}(z)$ is the fiducial model.     For details of the fiducial model, see Ref.~\cite{Heinrich:2016ojb} and note that
hydrogen is fully ionized for $z<z_{\rm min}$ whereas it follows the recombination ionization fraction for $z>z_{\rm max}$ and is constant in between.  We choose $z_{\rm max}=30$ since the \Planck data do not significantly
  prefer any ionization above this redshift \cite{Heinrich:2018btc}.
We retain the first 5 principal
components since they suffice to describe any ionization history in this range to the cosmic variance limit.

For arbitrary $t_a$, the ionization fraction described by Eq.~(\ref{eqn:reionPC}) can exceed the physical bounds imposed by zero
and full ionization.
We follow Refs.~\cite{Mortonson:2007hq,Heinrich:2016ojb} in placing necessary but not sufficient priors on $t_a$ to limit unphysical behavior.
We cannot impose strictly sufficient priors using only 5 principal components because the omitted components, although irrelevant for the observable power spectrum, do affect the
physicality of the ionization history.    These omitted components tend to give oscillatory and not cumulative contributions (see Fig.~\ref{fig:futureXe})  and
so we can better identify model-independent constraints on the ionization history through the cumulative
optical depth \begin{align}
  \tau(z,z_{\rm max}) = n_{\rm H} \sigma_T \int_z^{z_{\rm max}}dz\frac{x_e(z)(1+z)^2}{H(z)},
\end{align}
where $n_{\rm H}$ is the number density of hydrogen today and $\sigma_T$ is the  Thomson  cross section.
  Of course when using these PC constraints for testing specific physical models, physical priors are automatically established and  $t_a$ constraints can be applied directly \cite{Heinrich:2016ojb,Miranda:2016trf,Ahn:2012sb}.

\begin{table}
\begin{tabular}{|c|c|c|}
  \hline
  Model & Added Param. & $-2\Delta\ln{\mathcal{L}}$\\
  \hline \hline
  $\Lambda$CDM & $\tau$ & 0.0\\
  Rei & $t_a$ & 5.7\\
  Ifn & $\tau, p_i$ & 17.9 \\
  Ifn+Rei & $t_a, p_i$ & 22.3 \\
  \hline
\end{tabular}
  \caption{\footnotesize
 Models, the parameters they add to  the fundamental set
 $\{\ln A_s, n_s, \theta_{\rm MC}, \Omega_bh^2, \Omega_ch^2 \}$, and their maximum likelihoods
 relative to $\Lambda$CDM.   In $\Lambda$CDM and Ifn, reionization is parametrized by the total optical depth $\tau$ of a steplike transition.
  The 20 parameters $p_i$ are spline basis coefficients that generalize the inflationary tilt on large scales \cite{Obied:2017tpd}. The  5 parameters $t_a$ are the amplitudes of ionization history principal components \cite{Heinrich:2016ojb}.
 }
   \label{tab:models}
\end{table}

On the other hand, for the inflationary curvature  spectrum
we parameterize the local slope of the power spectrum in a manner consistent with
inflationary dynamics using the Generalized Slow-Roll (GSR) formalism \cite{Dvorkin:2009ne,Dvorkin:2010dn,Dvorkin:2011ui,Hu:2011vr},
\begin{align}
  \frac{d\ln\Delta_\mathcal{R}^2}{d\ln k} \rightarrow -G' \equiv (n_s-1)  -\delta G',
  \label{eqn:Gprimetilt}
\end{align}
with
\begin{align}
 \delta G'(\ln s) = \sum_i p_i B_i(\ln s).
\end{align}
Deviations from a constant slope are characterized by amplitudes $p_i$ and a spline basis $B_i(\ln s)$
where $s \equiv \int d \ln a\,\, c_s(aH)^{-1}$ is the inflaton sound horizon.
Unlike a direct parametrization of $\Delta_{\curv}^2$, this technique has the benefit of
automatically enforcing the inflationary requirement that the sharper the temporal feature, the more it
rings to higher $k$, which is important for modeling sharp features in the $TT$ spectrum.
 We choose
20 logarithmically spaced spline knots in the range $200 < s/{\rm Mpc} < 20000$ in order to
parameterize large-scale deviations from power-law initial conditions.
We restrict
 the range of $G'$  so as to bound power spectrum corrections that violate the
 linearity of Eq.~(\ref{eqn:Gprimetilt})  by taking the
second order parameter $|I_1|< 1/\sqrt{2}$, as discussed in Ref.~\cite{Dvorkin:2011ui}.

 In order to eliminate correlations between the amplitudes $p_i$ and identify the strongest constraints, we then define the inflationary principal component amplitudes $m_a \equiv \sum_i p_i V_{ia} $ where $V_{ia}$ is an orthonormal matrix of eigenvectors of the covariance matrix between the parameters $p_i$ obtained from the MCMC chains. We choose to report the 3 $m_a$ coefficients with the smallest errors.  Note that unlike reionization PCs, which are precomputed for completeness  from a cosmic variance limited measurement of the
 fiducial model, the inflationary PCs are defined with respect to the \Planck data.  When adding reionization freedom to inflationary freedom, we keep the principal component basis
 $V_{ia}$ fixed to those determined for the steplike reionization transition in order to enable direct comparisons between the
 two cases.
 For visualization purposes we construct the 3 PC filtered deviations from power
 law spectra
 \begin{equation}
  \delta G'(\ln s_i)  = \sum_{a=1}^3 m_a V_{ia}
  \end{equation}
  and spline interpolate between the $i$ samples.  Likewise we construct the 3 PC filtered
  version of the curvature power spectrum $\Delta_{\cal R}^2$  from $\delta G'$.   This visualization highlights the impact of the
   most significant aspects of the deviations but should not be interpreted as direct constraints on $\Delta_{\cal R}^2$
   (see Fig.~\ref{fig:Delta2full}).

  In summary, the models used in this paper  correspond to replacing the
optical depth $\tau$ of a steplike ionization history with the reionization parameters $t_a$ and/or adding the inflationary parameters $p_i$ to the fundamental $\Lambda$CDM parameters $\{\Omega_bh^2,\Omega_ch^2, \theta_{\rm MC}, \ln A_s, n_s\}$ (see Tab.~\ref{tab:models}).   We assume flat priors in each of these parameters out to ranges that are either
uninformative compared with the data or limited by the above-mentioned
restrictions on reionization and inflation.
Note that for reionization, a flat prior in $t_a$ does not correspond to
a flat prior in the total $\tau$ due to the enhanced freedom to vary the ionization history
at high redshift \cite{Mortonson:2008rx}.  Any preference for higher $\tau$ should be accompanied by a better fit to the data or interpreted in a model context with priors on physical parameters (e.g.~\cite{Miranda:2016trf,Heinrich:2018btc}).

\begin{figure*}
\centering
\includegraphics[width=0.85\textwidth]{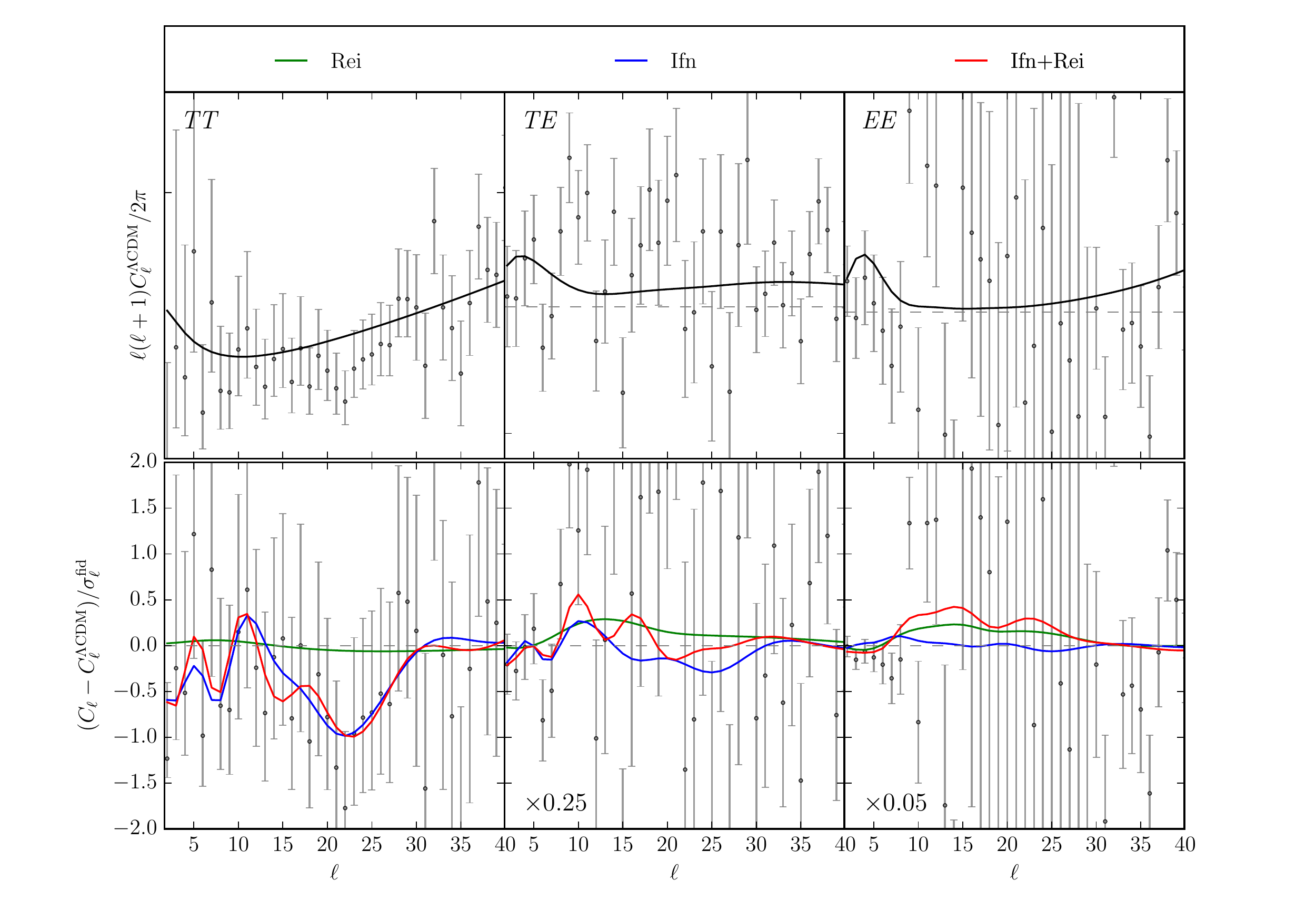}
\caption{\footnotesize \Planck power spectrum data vs.~the best fit models within the various classes.   Top panel: baseline $\Lambda$CDM model with power law inflationary curvature spectrum and steplike reionization.   Bottom panel: residuals with respect to $\Lambda$CDM in the data and the models with additional freedom in reionization (Rei), inflation (Inf), and both
(Inf+Rei). Residuals are scaled to the cosmic variance per $\ell$ of the fiducial reionization model as well as a further
scaling of $0.25$ ($TE$) and $0.05$ ($EE$) to fit to the $TT$ scale.
  Features in $TT$ drive inflationary constraints, especially around $\ell \sim 20$ and
those in $EE$ drive reionization constraints, especially around $\ell \sim 10$.
}
\label{fig:MLModels}
\end{figure*}

\begin{figure}
\centering
\includegraphics[width=0.45\textwidth]{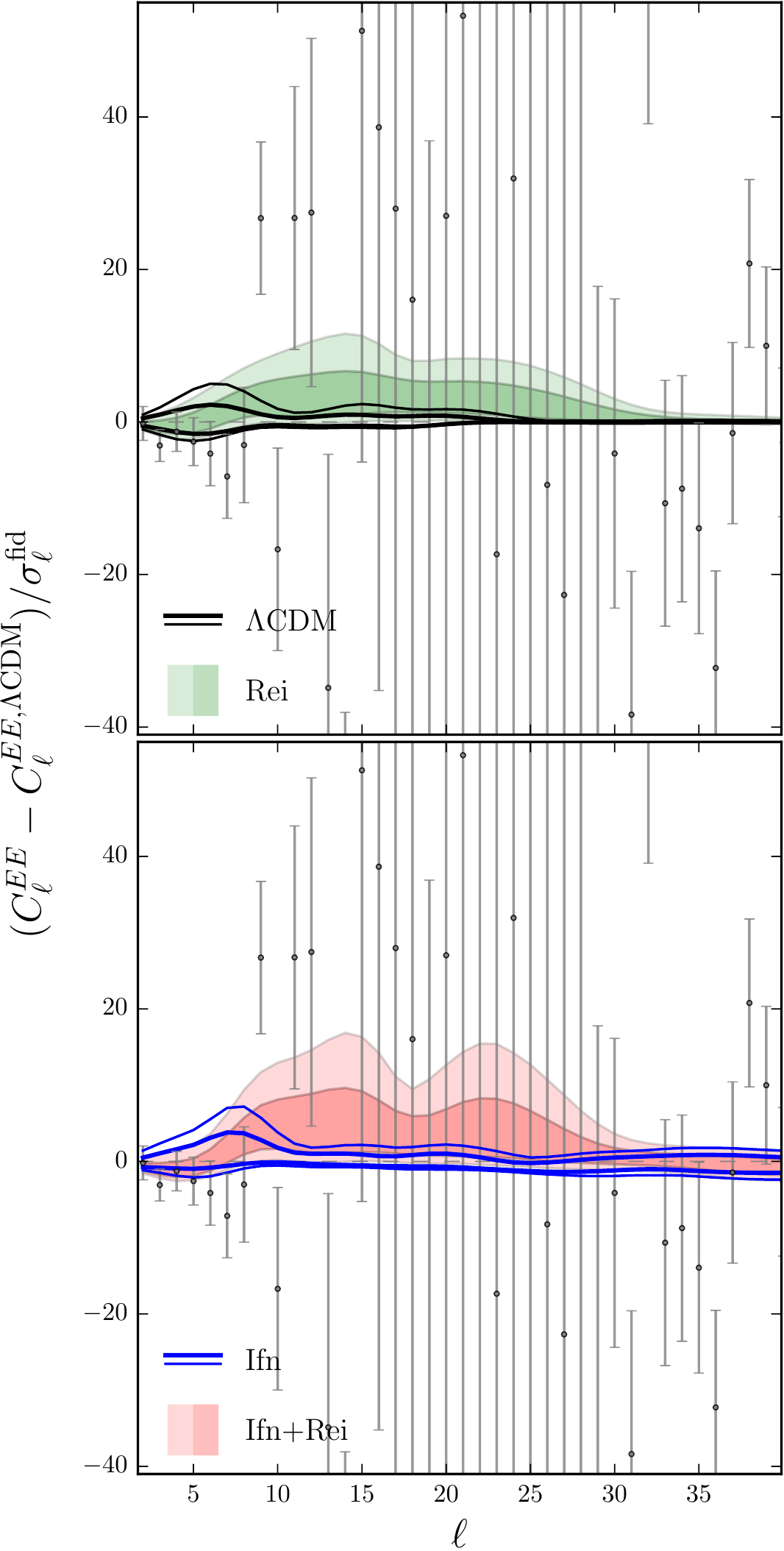}
\caption{\footnotesize
$EE$ power spectrum constraints for the various model classes (68\% and 95\% CL).  Constraints are driven by the high power at $\ell=9$ and the several following multipoles
which favor high redshift ionization in the Rei and Inf+Rei classes and cannot be fit with
Inf alone.
Models and data are plotted with respect
to the best fit $\Lambda$CDM model as in Fig.~\ref{fig:MLModels}.}
\label{fig:ClEElin}
\end{figure}

\begin{figure}
\centering
\includegraphics[width=0.49\textwidth]{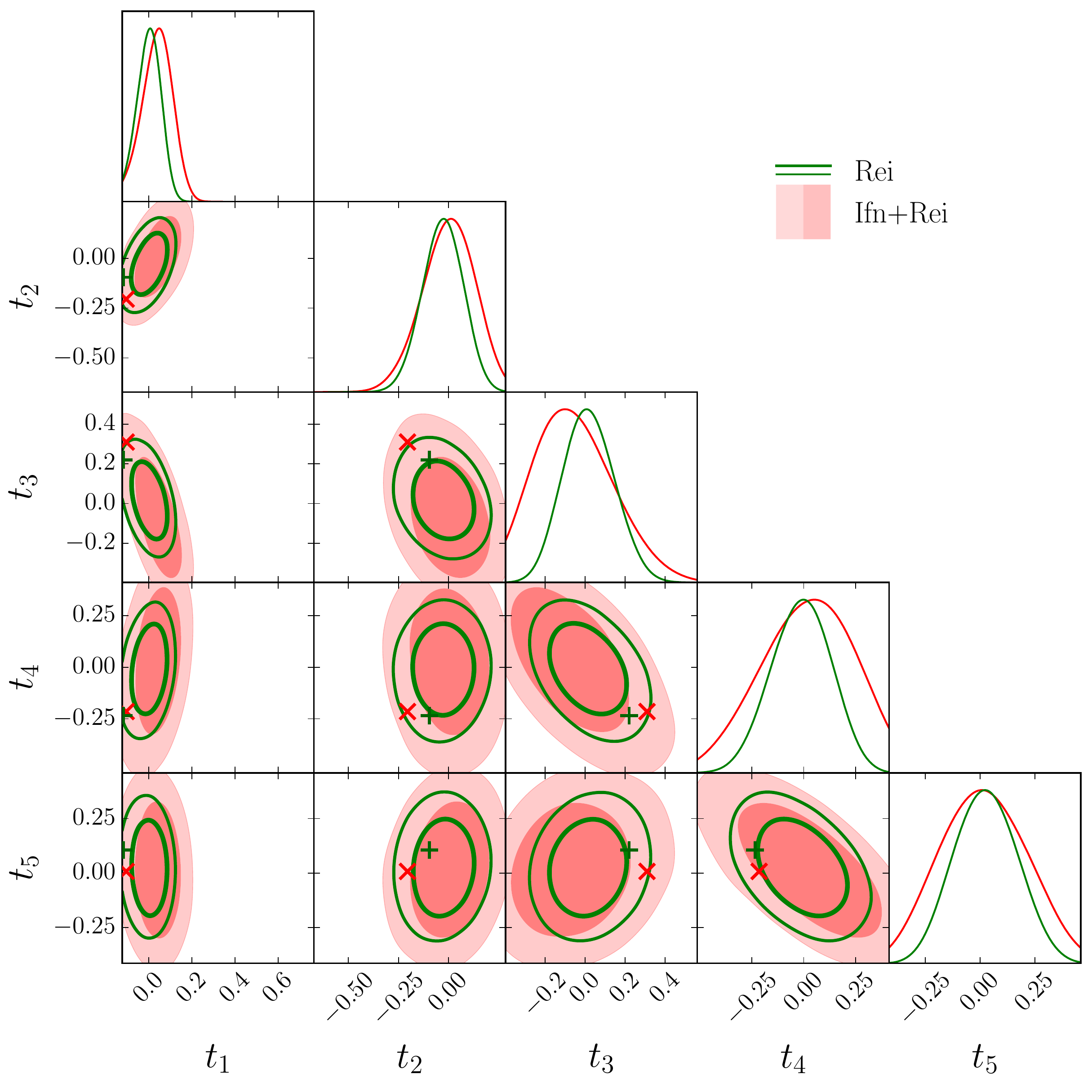}
\caption{\footnotesize Reionization PC constraints with (red) and without  (green)
inflationary $p_i$ parameters marginalized (68\% and 95\% CL).
The red $\times$ and green $+$ indicate the best fit steplike reionization history in the
respective classes and lies outside the preferred regime, especially in $t_1$ and $t_2$. }
\label{fig:reiPCtriangle}
\end{figure}

The maximum likelihood (ML) model of each parametrization is shown along with the data in Fig.~\ref{fig:MLModels}.
In the lower panels we scale the residuals against the $\Lambda$CDM ML model
to the cosmic variance per $\ell$ mode
 \begin{equation}
 \sigma_{\ell} =
 \begin{cases}
\sqrt{\frac{2}{2\ell+1}} C_\ell^{TT}, & TT;\\
\sqrt{\frac{1}{2\ell+1}}  \sqrt{ C_\ell^{TT} C_\ell^{EE} + (C_\ell^{TE})^2}, & TE;\\
\sqrt{\frac{2}{2\ell+1}}  C_\ell^{EE}, & EE, \\
\end{cases}
\end{equation}
evaluated using the fiducial reionization model.   We use this model for convenience since  the $\Lambda$CDM steplike reionization history produces an $EE$ spectrum whose cosmic variance
is too far below the \Planck noise variance at $10 \lesssim \ell \lesssim 30$ to be practical.
Note that the data and models in Fig.~\ref{fig:MLModels} have been rescaled by $1/4$ for $TE$ and
$1/20$ for $EE$ to fit on the same scale as the near cosmic variance limited $TT$ measurements.   This scaling also highlights that there is significant
opportunity for improvement in polarization measurements which can be used to test the
interpretation of
both reionization and inflation features in the future.

The likelihood improvements of these models over $\Lambda$CDM  is given in Tab.~\ref{tab:models}.   While these improvements do not justify the introduction of 5 ionization parameters and 20 new inflationary parameters, our parameterization aims for completeness
in the characterization of power spectra features.   These model-independent constraints can then be used to test specific models
with fewer parameters or
find matching features within and
between the temperature and polarization spectra that could reveal their physical origin.

 For reionization, the 5 parameters currently represent one new aspect of the model class, the ability to have a significant component of the total optical depth at $z>10$
\cite{Heinrich:2018btc}, whereas for inflation the 3 PC compression highlights the coherent features rather than the statistical fluctuations of the low-$\ell$ power spectra.  Notice also that the improvements from reionization and inflation parameters are nearly additive, indicating that
they are controlled by almost independent aspects of the data.

Indeed the reionization and inflation parameters fit mostly separate features in the
various  power spectra at low-$\ell$ as shown in Fig.~\ref{fig:MLModels}.    The well  known deficit of power in $TT$
around $\ell\sim 20$  mainly drives the inflationary degrees of freedom whereas the excess
in $EE$ around $\ell \sim 10$ drives preferences for high redshift ionization.  The $TE$ power spectrum naturally combines features in both $T$ and $E$ due to their correlation.
We explore next the interplay between these features in the data and constraints on
reionization and inflation.

\section{Reionization and Inflation}
\label{sec:rei_infl}

In this section, we discuss the implications of polarization and temperature power spectra features on separate and joint constraints of reionization history and inflation.

\begin{figure}
\centering
\includegraphics[width=0.45\textwidth]{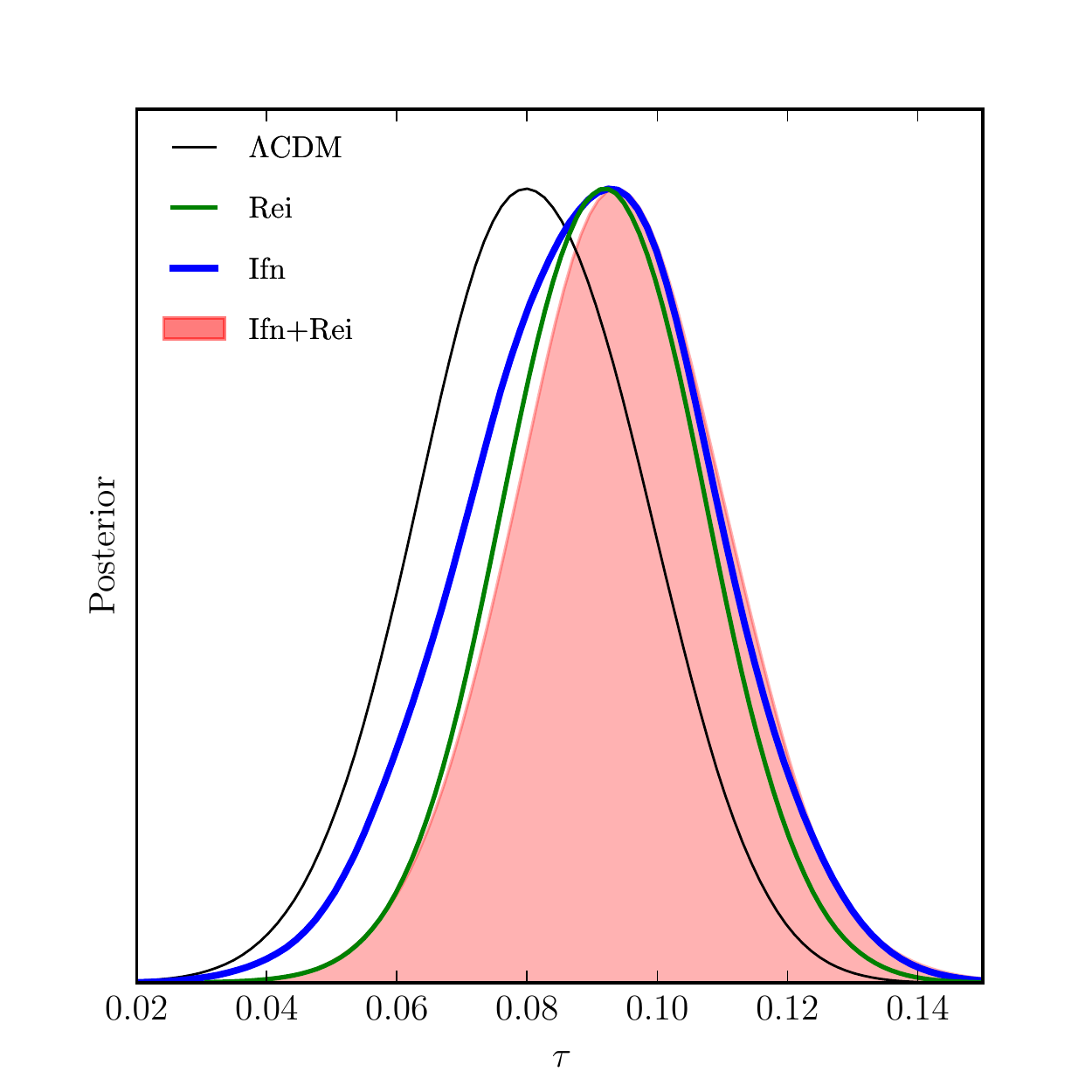}
\caption{\footnotesize Total optical depth $\tau$ constraints in the
various model classes. }
\label{fig:taushift}
\end{figure}

\subsection{Polarization Features and Reionization History}

CMB polarization is particularly sensitive to the ionization history since it is generated only when a
quadrupole temperature anisotropy is scattered by free electrons.
During reionization, the
$EE$-mode polarization spectrum gains a feature at scales that roughly correspond to the horizon size at the respective redshift where these quadrupoles form from streaming radiation.
 Measurements of this feature
provides coarse-grained constraints on the ionization history, and in particular, on the amount of high vs.~low $z$ optical depth.

On the other hand, the usual imposition of a steplike reionization requires the optical depth to
mainly come from low redshifts.
   In the Rei model we relax this assumption by adding the 5
PCs $t_a$. In the Rei+Ifn model, we test the robustness of reionization constraints to features in the curvature power spectrum.

Ref.~\cite{Heinrich:2016ojb} found that once the steplike imposition is relaxed with reionization PCs, the \Planck 2015 data not only allows  but also is better fit by a high $z \gtrsim 10$ component with a 95\% CL preference for finite contributions at $z\gtrsim 15$ (see also \cite{Hazra:2017gtx,Heinrich:2018btc}).  In Fig.~\ref{fig:MLModels} we see that this preference is driven by
a sharp increase in $EE$ power at $\ell =9$ with several points that average high thereafter.
    As pointed out in Ref.~\cite{Aghanim:2015xee},  $\ell=9$  is anomalously high at 2.7$\sigma$ if a steplike reionization is assumed.  In addition $\ell=9$ is slightly low in $TT$ even though the two should be positively correlated.   In the steplike model, power at these multipoles cannot be raised without violating the constraints from the very low power at $2 \le \ell \le 8$.
    In fact, in the steplike model, the best fit is a compromise between these very low and very high points as shown in Fig.~\ref{fig:MLModels}.
    By allowing low ionization at low redshift and high ionization at high redshift relative to a step, the reionization PCs
    can better thread through these constraints.
These models make very different and testable predictions for the polarization spectrum at $10 \lesssim \ell \lesssim 40$.

 In Fig.~\ref{fig:ClEElin} we show the posterior distribution of $C_\ell^{EE}$ from the various cases analyzed in this paper. The ability to raise $EE$ power at $9 \le \ell \le 15$ requires reionization freedom as can be seen from the similarity of Rei and Inf+Rei and conversely the lack of power in {both} $\Lambda$CDM and Inf constraints.   Inflationary degrees of freedom  produce a matching set of features between $EE$ and $TT$ once projection effects are taken into account.  Therefore they cannot alone be responsible for a feature in $EE$ that is not present in $TT$.

  When added to reionization features, inflationary features  can marginally help sharpen the rise in power at $\ell\sim 9$ due to a matching rise  at $\ell \sim 11$ in $TT$.  Note also that cosmic variance  does not
 correlate statistical fluctuations at different multipoles between the two spectra despite the $TE$ correlation, whereas physical effects require an offset in multipoles
 due to projection effects \cite{Mortonson:2009qv}. Cosmic variance does make the significance of this joint feature
 weak.  With Ifn+Rei, the data favor slightly less ionization at $z \lesssim 15$ and allow more at $z \gtrsim 15$ which then leads to
more freedom to raise $EE$ in the $20 \lesssim \ell \lesssim 25$ regime where a signal is  yet to be measured.  This
freedom can fully mask features that come  from inflation alone shown in
Fig.~\ref{fig:ClEElin} which are
due to the $TT$ feature in this regime
 unless they can be excluded by measurements in other $\ell$ ranges  \cite{Mortonson:2009qv} or in the context of specific physical reionization models.

\begin{figure}
\centering
\includegraphics[width=0.45\textwidth]{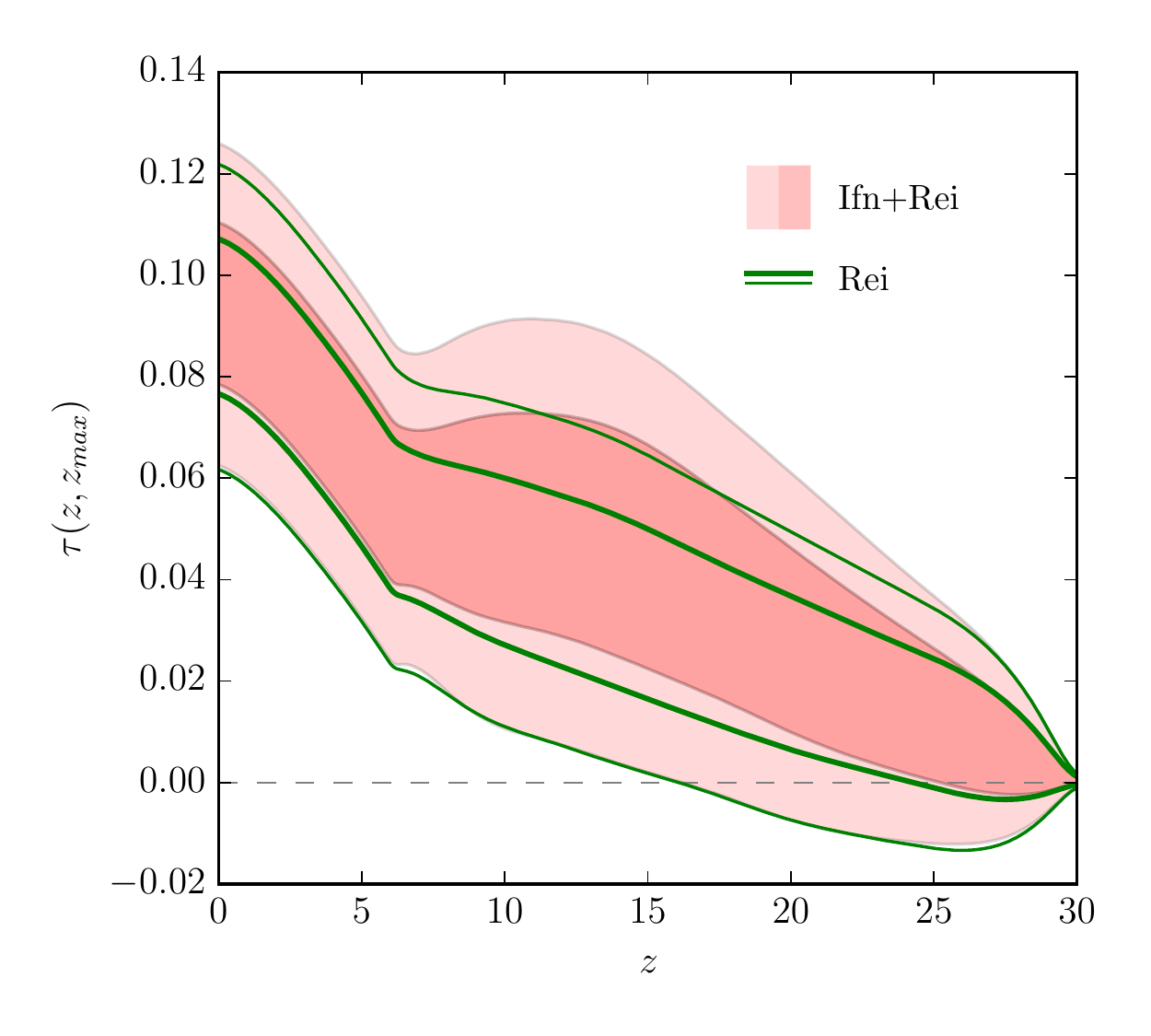}
\caption{\footnotesize Cumulative $\tau$ constraints (68\% and 95\% CL) for the
Rei and Inf+Rei class.   In both cases there is a 95\% CL preference for ionization at
$z\gtrsim 15$ that cannot be accommodated with a steplike reionization history. }
\label{fig:TauZ}
\end{figure}

We can also see these effects in the constraints on the reionization PCs $t_a$ and their implications for the cumulative
optical depth in Figs.~\ref{fig:reiPCtriangle} and \ref{fig:TauZ}.  Constraints on $t_a$ broaden moderately when marginalizing
the additional $p_i$ inflationary degrees of freedom.   The boundaries of the panels represent the weak physicality constraint
on the ionization history described in the previous section.
Without the inflationary degrees of freedom, the priors just
clip the allowed region for negative values of $t_1$, corresponding to forbidding negative ionization at $z \gtrsim 15$.
The fact that the posterior peaks away from this boundary indicates that the data in fact favor some ionization at $z \gtrsim 15.$
This is borne out by the cumulative $\tau$ constraints where there is a 95\% CL preference for finite contributions there.
Were it not for the priors, the impact of marginalizing $p_i$ on $t_4$ and $t_5$ would be more apparent as these parameters are
currently better constrained by the reionization Doppler effect on the temperature power spectrum at intermediate multipoles below the
first acoustic peak than by the polarization power spectrum at low multipoles.

With inflationary degrees of freedom, physicality bounds
 also clip positive fluctuations in $t_2$ and negative fluctuations in $t_3$ which
correspond to forbidding negative ionization at $z \lesssim 15$.     In this case, models with essentially no ionization at
$6 \lesssim z \lesssim 15$ are allowed.
 In fact the weak physicality priors still allow some
negative ionization as discussed above.   This can be seen in Fig.~\ref{fig:TauZ} where,
within the 95\% CL bounds, the cumulative optical depth is allowed to be non-monotonic with a peak between $10 \lesssim z \lesssim 15$.   These reionization constraints for Rei+Ifn therefore err on the conservative side, especially in allowing additional
excess $EE$ power at $20 \lesssim \ell \lesssim 25$ over Rei.

Nonetheless, in the other direction, the 95\% CL preference for finite contributions to $\tau$ at $z\gtrsim 15$ remains robust
to inflationary features.  This corresponds to the fact that inflationary freedom alone cannot substantially raise the
$EE$ power at $\ell \ge 9$ without changing the ionization history or violating $TT$ constraints.
Correspondingly in Fig.~\ref{fig:reiPCtriangle} the best fit steplike reionization model with ($\times$) and without
($+$) marginalizing inflationary parameters lies in the disfavored region especially in the
$t_1-t_2$ plane.

The $z=0$ endpoint of the cumulative $\tau$ constraint is the total optical depth. In Fig.~\ref{fig:taushift}, we display its
posterior in the various cases. In moving from $\Lambda$CDM to Ifn, $\tau$ increases but its errors remain similar.  This is
due to the fact that features in $TT$ favor a net suppression of inflationary power so, in order to achieve the same $EE$ power, we
require higher $\tau$.   In the Rei model, $\tau$ increases and the width decreases.
 This reflects the ability to relax the tension  that occurs in the steplike model between the low $EE$ power
at $\ell \le 8$ and the high $EE$ power at $\ell > 8$.
Moving to the Ifn+Rei model has little impact on $\tau$ since the enhanced freedom to
have ionization at $z \gtrsim 15$ is compensated by lower ionization at lower redshifts.
Note also that the upper bounds on $\tau$ remain fairly robust in all of the extensions.  The low power at $\ell\le 8$ constrains
all models since high redshift ionization unavoidably contributes there as well due to projection effects
(see  Fig. 2 in Ref.~\cite{Hu:2003gh}).

\begin{figure}
\centering
\includegraphics[width=0.45\textwidth]{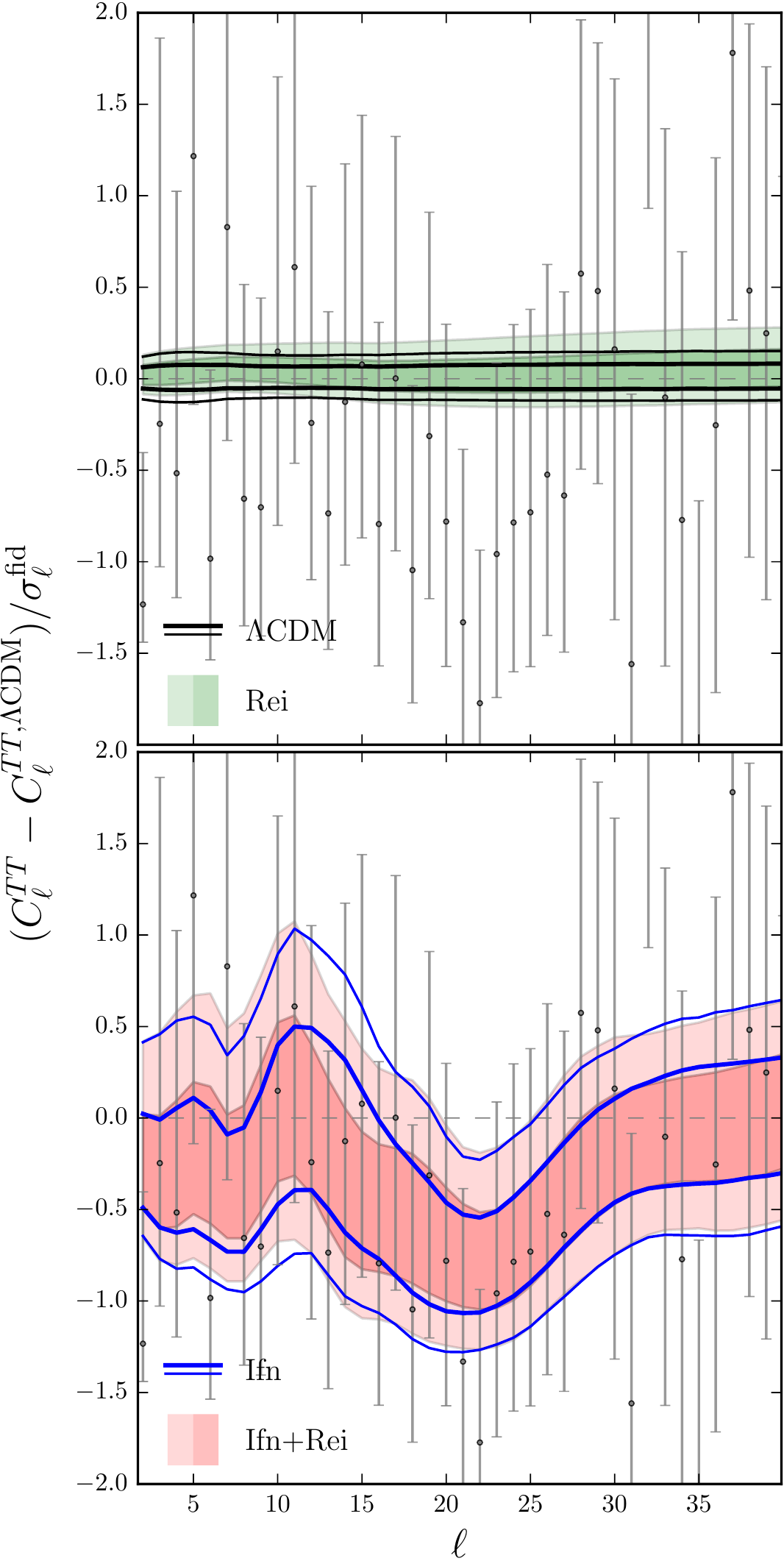}
\caption{\footnotesize $TT$ power spectrum constraints for the various model classes (68\% and 95\% CL).   Constraints are driven by the low power glitch around $\ell \sim 20$ which can be fit by models with inflationary freedom Inf and Inf+Rei but not by Rei alone.
Models and data are plotted with respect
to the best fit $\Lambda$CDM model as in Fig.~\ref{fig:MLModels}}
\label{fig:ClTTlin}
\end{figure}

\begin{figure}
\centering
\includegraphics[width=0.45\textwidth]{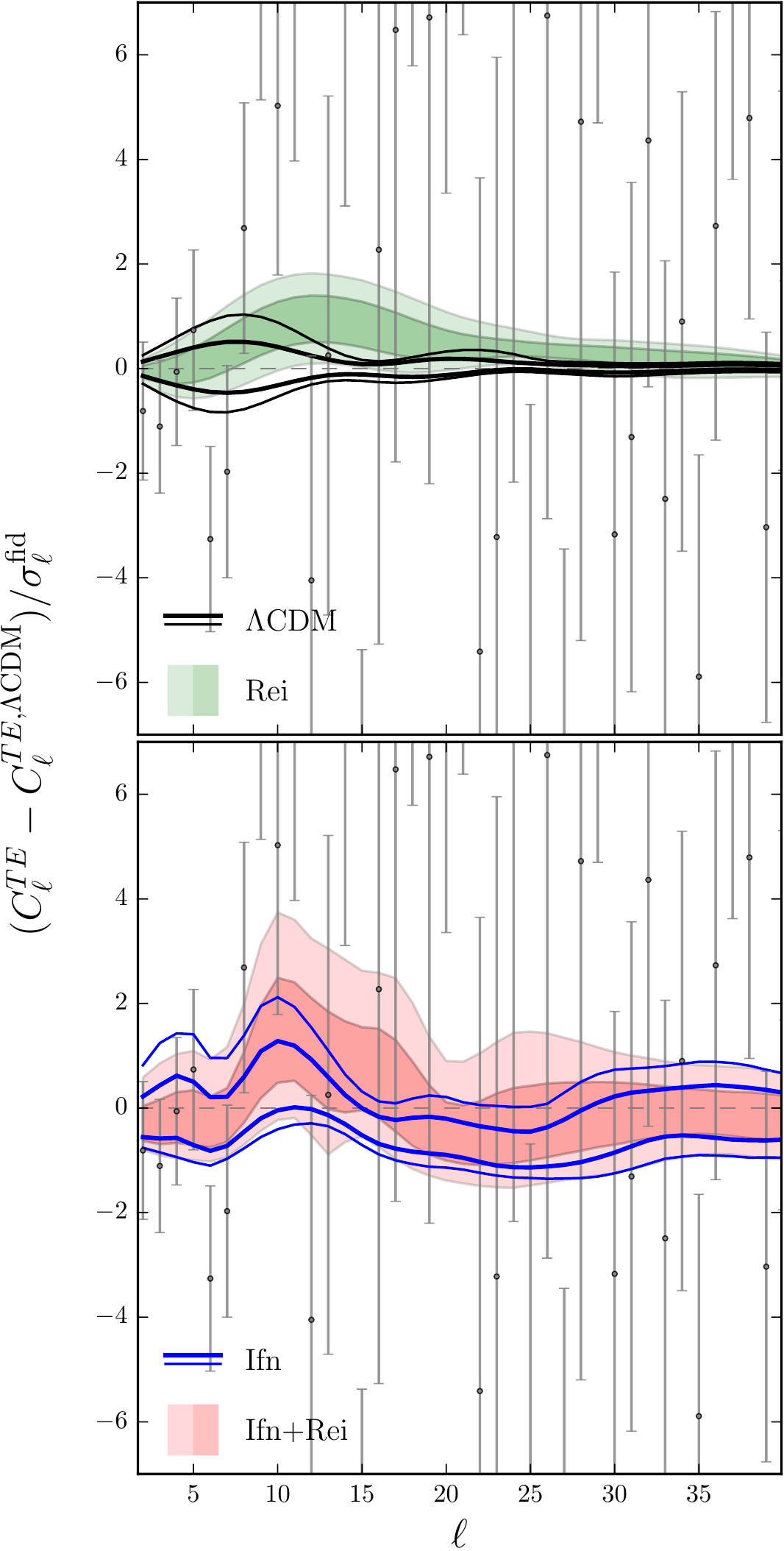}
\caption{\footnotesize $TE$ power spectrum constraints for the various model classes (68\% and 95\% CL).   Constraints follow the combined features of $EE$ and $TE$ shown in
Figs.~\ref{fig:ClEElin} and \ref{fig:ClTTlin} allowing the Inf+Rei to best fit the  $10 \lesssim \ell \lesssim 20$ regime.
 Models and data are plotted respect
to the best fit $\Lambda$CDM model as in Fig.~\ref{fig:MLModels}. }
\label{fig:ClTElin}
\end{figure}

\subsection{Temperature Features and Inflation History}

The well known anomalies in the low multipole $TT$ spectrum drive the constraints on features
during inflation.
In Fig.~\ref{fig:ClTTlin}, we show the posterior constraints on the $TT$ spectrum in the
various model classes.  In particular both Ifn and Ifn+Rei attempt to fit the suppression of
power at $\ell \sim 20$ and a rise at $\ell \sim 11$ using the inflationary freedom, whereas neither
$\Lambda$CDM nor Rei have the ability to do so.  In these cases, the fiducial constant tilt model
lies outside the 95\% CL band from $20 \le \ell \le 25$.

As discussed in the previous section, the
main impact of Rei freedom on the $TT$ posterior is to slightly sharpen the rise at $\ell \sim 11$ due
to the matching feature in polarization at $\ell \sim 9$ and the reduced necessity of excess curvature power to explain the $EE$ spectrum at higher multipoles.  While with inflationary parameters alone, Figs.~\ref{fig:MLModels} and \ref{fig:ClEElin} show that the $TT$ feature
at $\ell \sim 20$ is matched by an $EE$ feature
at $\ell \sim 25$  that is larger than the cosmic variance errors, they are much smaller than the \Planck polarization errors there and can also
be masked by reionization features.

The impact on the $TE$ posterior from Inf+Rei is stronger, as shown in Fig.~\ref{fig:ClTElin}.
$TE$ represents a combination of temperature and polarization features.   Ifn alone produces
similar features in $TE$ as in $TT$.   Rei alone produces a smooth excess in $TE$ power at $\ell \sim 10-15$.
The Inf+Rei combination then contains elements of both showing a broad but pronounced feature at $\ell \sim 10-15$, which best match the data.

\begin{figure}
\centering
\includegraphics[width=0.45\textwidth]{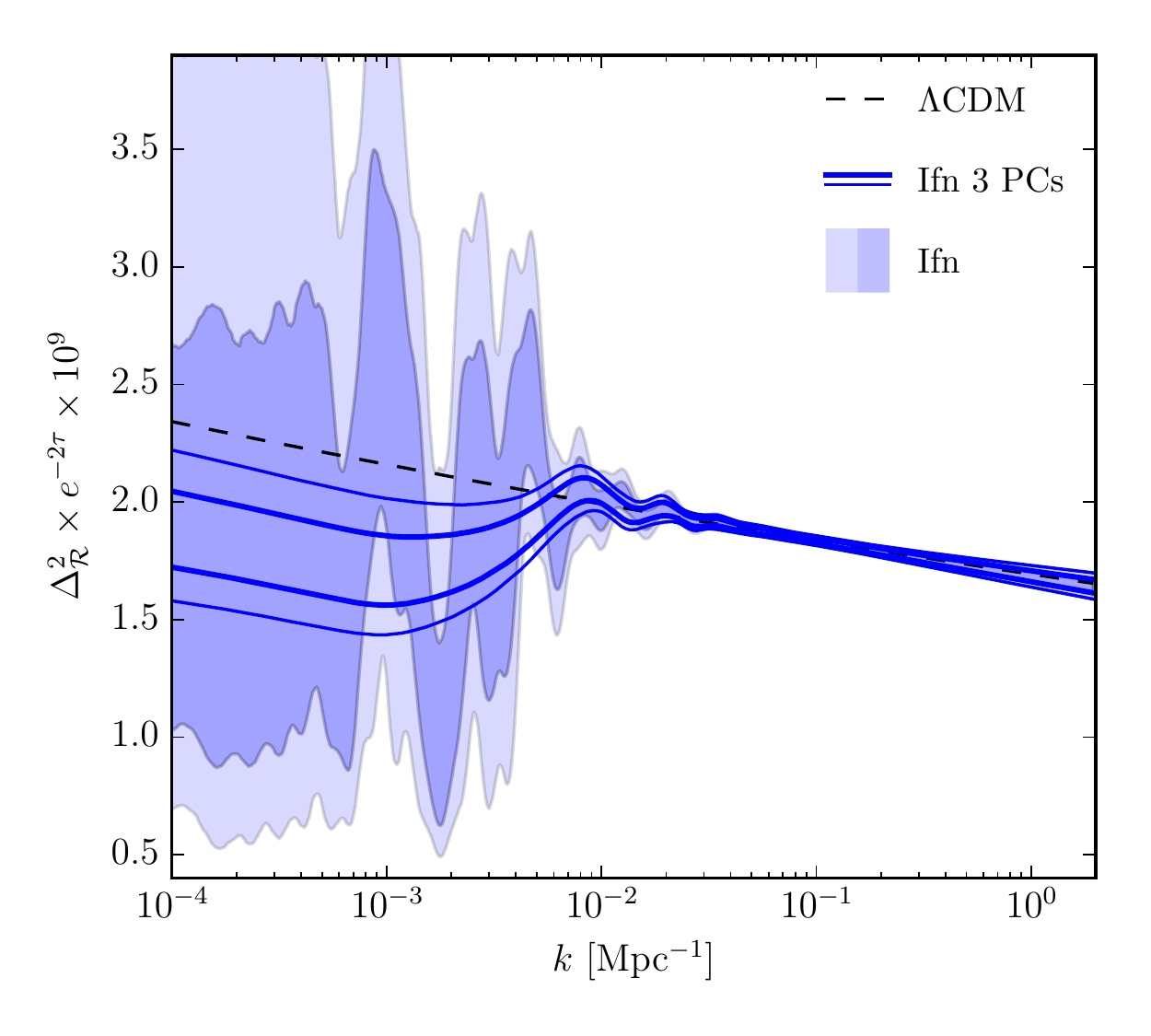}
\caption{ \footnotesize  Curvature power spectrum constraints for the Inf class (68\% and 95\% CL).  The 20 parameters $p_i$ fit fluctuations in the $TT$ power spectrum shown in Fig.~\ref{fig:ClTTlin} and imply constraints on $\Delta_{\curv}^2$ with the glitch at $\ell \sim 20$ corresponding to a dip at $k \sim 0.002$ Mpc$^{-1}$.  The 3 PCs $m_a$ filter out low significance features leaving a preference for less low $k$ power
than can be accommodated by slow-roll inflation.
}
\label{fig:Delta2full}
\end{figure}

Next we examine the implications of the \Planck data for the inflationary parameters and initial conditions.
In Fig.~\ref{fig:Delta2full} shows the posterior constraints on the curvature power spectrum in the Inf case.   Note that the
$TT$ feature at $\ell \sim 20$ corresponds to a dip in power at $k \approx 0.002$ Mpc$^{-1}$ whereas the rise at $\ell \sim 11$ to
the bump at $k \approx 0.001$ Mpc$^{-1}$.   By constructing  the curvature power spectrum $\Delta_\curv^2$ consistently from
the inflationary source $G'$, we enforce the requirement that sharp temporal features during inflation lead to oscillatory
features in the power spectrum.

\begin{figure}
\centering
\includegraphics[width=0.49\textwidth]{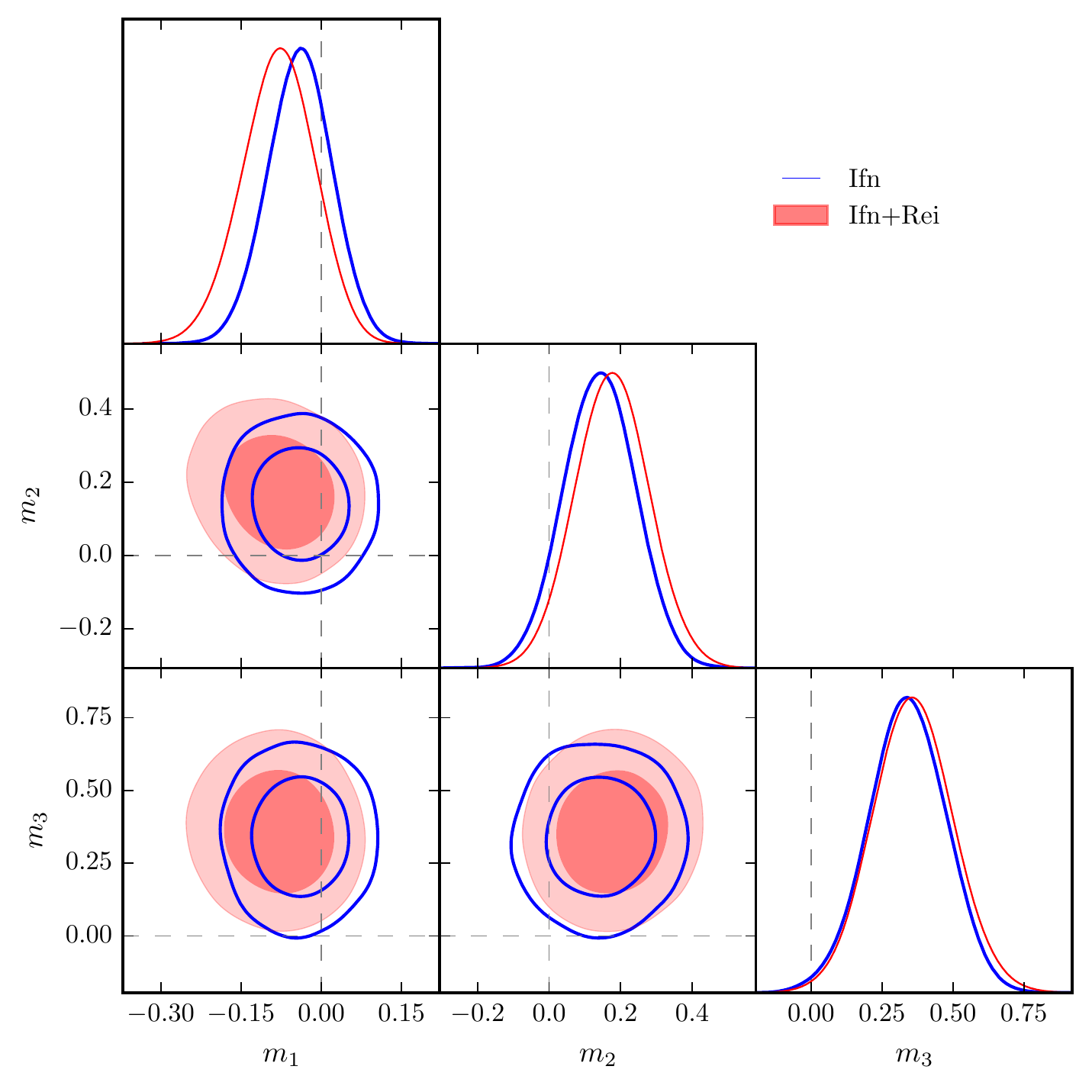}
\caption{\footnotesize Inflation 3 PC constraints (68\% and 95\% CL).   The preference for
deviations from power law inflationary conditions (dashed lines) remains and marginally increases once reionization parameters are marginalized in Inf+Rei.}
\label{fig:ifnPCtriangle}
\end{figure}

\begin{figure}
\centering
\includegraphics[width=0.45\textwidth]{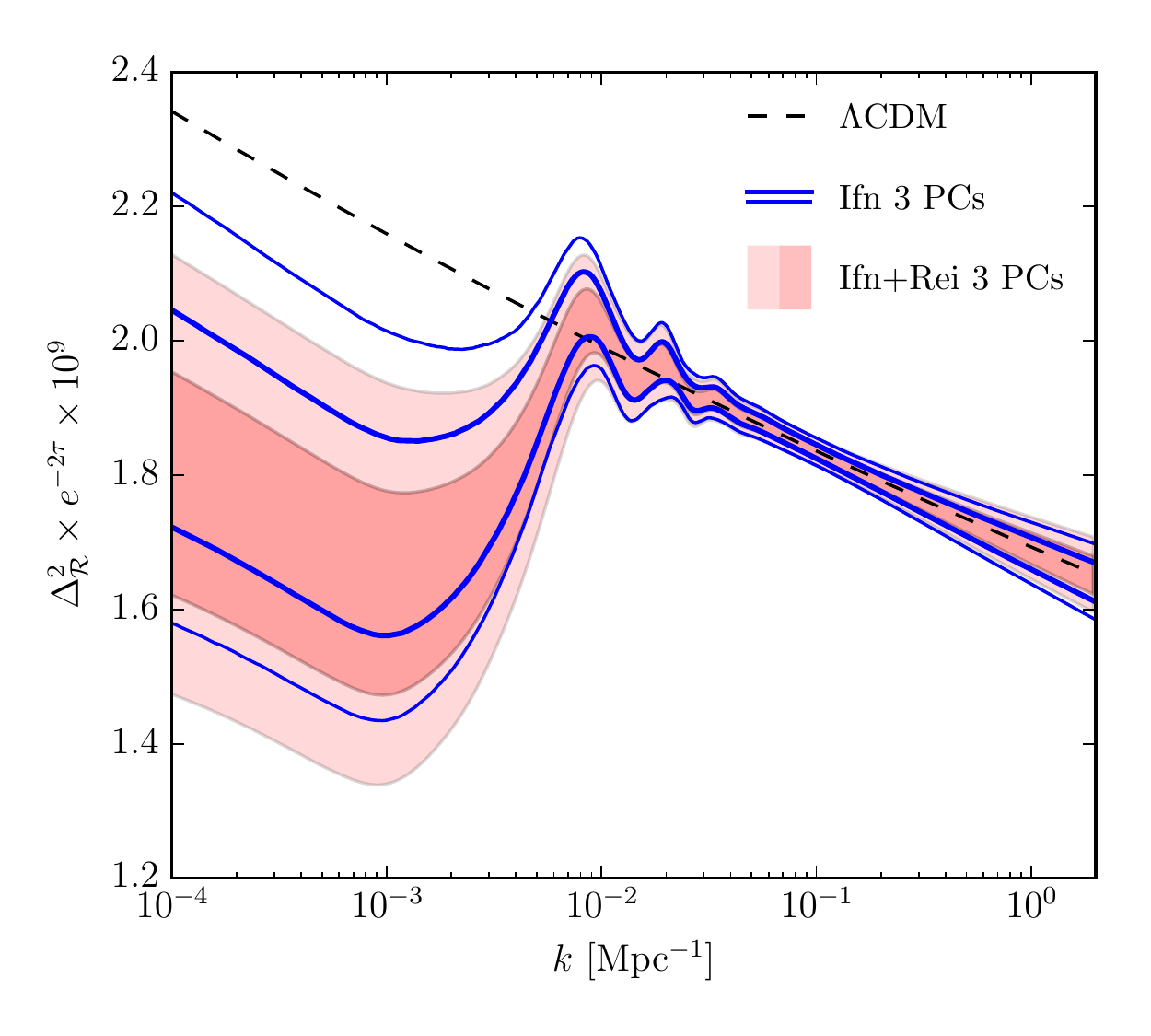}
\caption{ \footnotesize Inflation  3 PC impact on the curvature power spectrum in Inf+Rei compared with Ifn (68\% and 95\% CL).
  Marginalizing reionization parameters under Inf+Rei causes the net reduction of low $k$ power to become larger to compensate for the larger total optical depth $\tau$.
}
\label{fig:Delta2}
\end{figure}

On the other hand these oscillatory deviations also tend to fit out statistical
fluctuations in the data.    The inflationary PCs effectively filter out these low significance features.
In  Fig.~\ref{fig:ifnPCtriangle}, we show that inflationary principal component
constraints under Ifn.  Notice that the slow roll prediction of $m_1=m_2=m_3=0$ lies outside the 95\% CL region in the
$m_2-m_3$ plane.   In Fig.~\ref{fig:Delta2full} we show that this deviation is associated with a suppression in power
at $k \lesssim 0.005$ Mpc$^{-1}$.

\begin{figure}
\centering
\includegraphics[width=0.45\textwidth]{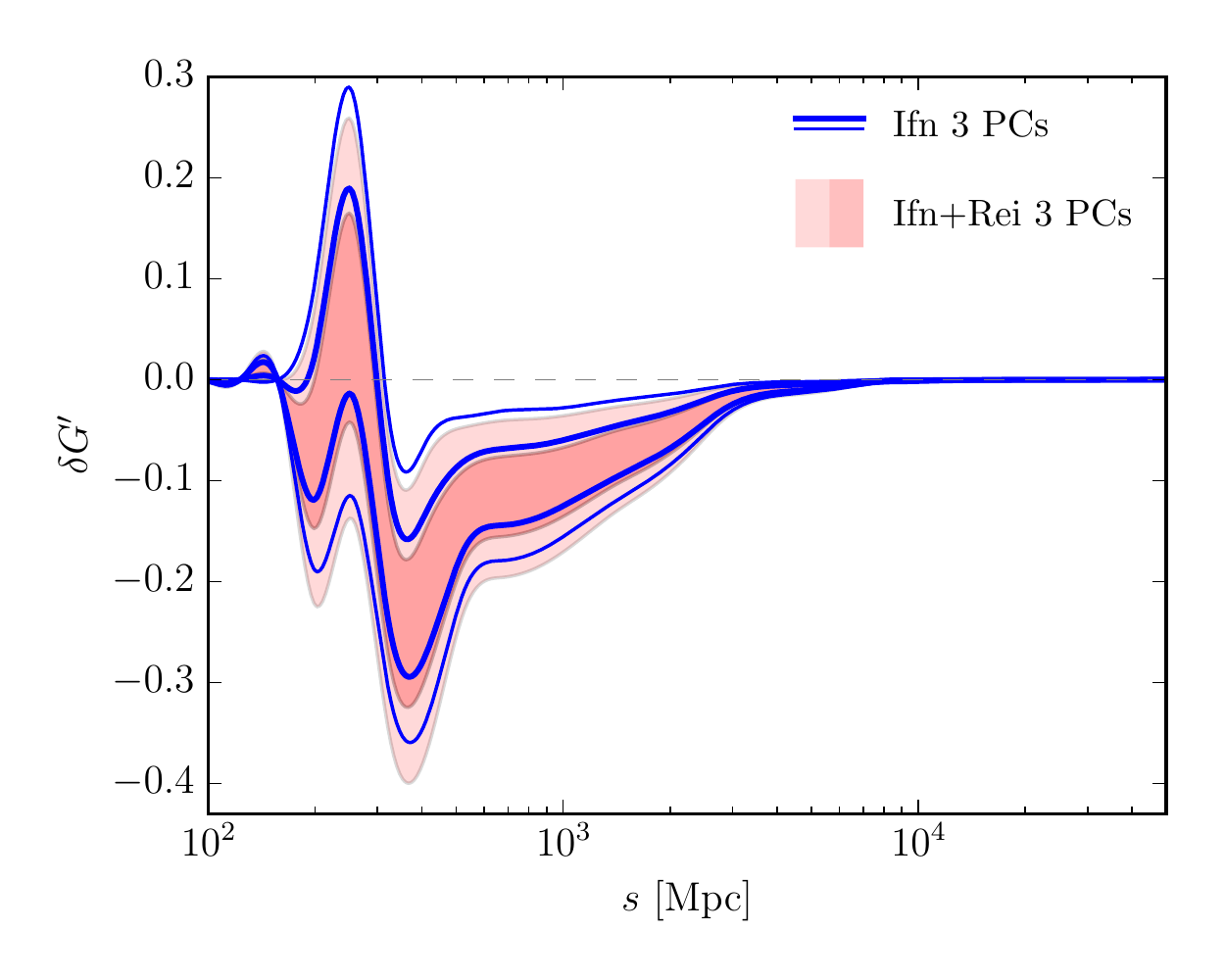}
\caption{ \footnotesize Inflation 3 PC impact  on  $\delta G'$, the source of deviations from constant tilt,
  (68\% and 95\% CL).  The feature at $\ell \sim 20$
in $TT$ produces a similar localization of a sharp feature during inflation when the sound
horizon was $s \sim 300$ Mpc.  Marginalizing reionization parameters in Inf+Rei makes the decrement slightly larger.}
\label{fig:Gprime}
\end{figure}

These conclusions
are largely robust to marginalizing reionization PCs in Inf+Rei.   In fact, counterintuitively all three PC
components deviate slightly more from zero in  Fig.~\ref{fig:ifnPCtriangle} once reionization is marginalized,
with little enhancement of their errors.
In Fig.~\ref{fig:Delta2} we show that correspondingly the suppression of
power at $k \lesssim 0.005$ Mpc$^{-1}$ is actually sharper once reionization is marginalized.   The localization is
even more apparent in the impact of the  3 inflationary PCs on $\delta G'$ shown in
Fig.~\ref{fig:Gprime}.  Again, the suppression at large
scales becomes slightly more rather than less significant when reionization is marginalized.

The reason for the enhancement of the inflationary feature by reionization can be understood mainly as the indirect effect
of the higher total optical depth $\tau$ favored in the Inf+Rei case shown in Fig.~\ref{fig:taushift}.
A larger $\tau$ lowers the acoustic peaks relative to the low-$\ell$ temperature power spectrum
and therefore requires a larger suppression of low to high $k$ from inflation to achieve the same $TT$ spectrum.   This larger suppression  comes from both a slightly larger tilt ($n_s = 0.962 \pm 0.005$ for Ifn and $n_s = 0.964 \pm 0.005$ for Ifn+Rei) and a larger feature from the
inflationary principal components.

These effects, however, are relatively minor so that the main conclusion is that the preference
for an inflationary feature to explain the  deficit of $TT$ power at $\ell \sim 20$ remains
even when all possible reionization histories between $6 \le z \le 30$ are marginalized
with the reionization PCs.

\section{Discussion and Conclusion}
\label{sec:conclusions}

The low order multipoles of the CMB temperature and polarization spectra show anomalous features that could be explained by  corresponding features during inflation and reionization. To the extent that these features overlap, a joint analysis of inflation and reionization could reveal degeneracies between the two or favor one type over the other.

We find that in the \Planck 2015 data the temperature and polarization features instead favor both
inflation and reionization features with little interference between the two.
 More specifically, residuals in the \Planck $EE$ power spectrum taken with respect to the best fit $\Lambda$CDM model show an excess in power around $\ell \sim 10$ with noise dominated measurements at $\ell \sim 20$, while the $TT$ power spectrum shows a suppression in power at $\ell \sim 20$ with no significant enhancement
at $\ell \sim 10$. It is thus not possible to have a model where inflation alone accounts for
the $\ell \sim 10$ feature nor is it possible for reionization to create a large
temperature feature at $\ell \sim 20$.

Beyond a steplike reionization history, high redshift reionization at $z\gtrsim 10$ can better
account for the high $EE$ power, especially at $\ell=9$, while simultaneously maintaining
low power at $\ell < 9$ that is required by the data.   Marginalizing inflationary parameters
does not reduce the significance of lower limits to the cumulative optical depth at
$z \gtrsim 10$.

Likewise, an inflationary feature that transiently violates the slow-roll approximation is compatible with the sharp suppression at $\ell \sim 20$.    The two sets of features do
interfere in the sense that  high redshift reionization could potentially mask the matching $EE$ feature predicted by an inflationary explanation for the $TT$ feature.
However, counterintuitively,
marginalizing over reionization PCs makes the inflationary feature slightly more
rather than less significant.   The larger optical depth associated with high redshift ionization
requires more suppression of large scale power relative to the acoustic peaks.

Physically, this interpretation of the two observed set of features implies  two independent events in the cosmic history, with those in the $TT$ power spectrum relating to effects of slow-roll violation during inflation while the ones in the $EE$ spectrum being the result of early reionization.  Currently the significance of these features is fairly low, with high redshift reionization
favored at $2\Delta\ln {\cal L} \sim 6$ and inflationary features separately at $2\Delta \ln {\cal L} \sim
18$ but with 20 extra parameters.  The 3 best constrained combinations of the 20 parameters are responsible for  sharply suppressing large scale inflationary power while the
rest optimize the fine scale features to  fluctuations in the $TT$ spectrum.

In the future, polarization measurements at $10 \lesssim \ell \lesssim 40$
could potentially improve by more than an order of magnitude before hitting the cosmic variance limit.
With such improvements, the inflation and reionization explanations of current measurements
can be more definitively tested and disentangled. In particular, as explicitly shown in  Appendix \ref{sec:appendix},  if the preferred ionization history lies in the center of the allowed region in Fig.~\ref{fig:ClEElin}, i.e. the Rei and Ifn+Rei contours, it could be distinguished from a $\Lambda$CDM model.

\appendix

\begin{table}[b]
\begin{tabular}{|c|c|}
  \hline
  Future Data & Label \\
  \hline \hline
  None & Current \\
  $\Lambda$CDM bf $\ell=14-30$& CV$_{\rm Tanh}$ \\
  Rei bf $\ell=14-30$ & CV$_{\rm PC}$ \\
  \hline
\end{tabular}
  \caption{A summary of the post-processed Rei chains, the new data added and our labelling conventions. For example, ``$\Lambda$CDM bf $\ell = 14-30$" means that CV-limited measurements for $C_\ell^{EE}$ from  $\ell=14-30$ identical in value to the {\it Planck} 2015 $\Lambda$CDM best fit were added to {\it Planck} 2015 $EE$ data.}
   \label{tab:futuremodels}
\end{table}

\section{Cosmic Variance Limited E-mode Measurements}
\label{sec:appendix}

In this Appendix, we demonstrate the ability of future CV-limited measurements of the $EE$ polarization spectrum to distinguish models with extended reionization from a steplike ionization history.  For simplicity, we focus on the Rei case, where the $EE$ spectrum cannot be altered by changes in the inflationary PCs.  We note that, since $\Delta C_\ell \propto C_\ell$ for a CV-limited measurement,  future $C_\ell$ constraints depend on the specific model
within the Rei class assumed for the projected measurements.

\begin{figure}
\centering
\includegraphics[width=0.45\textwidth]{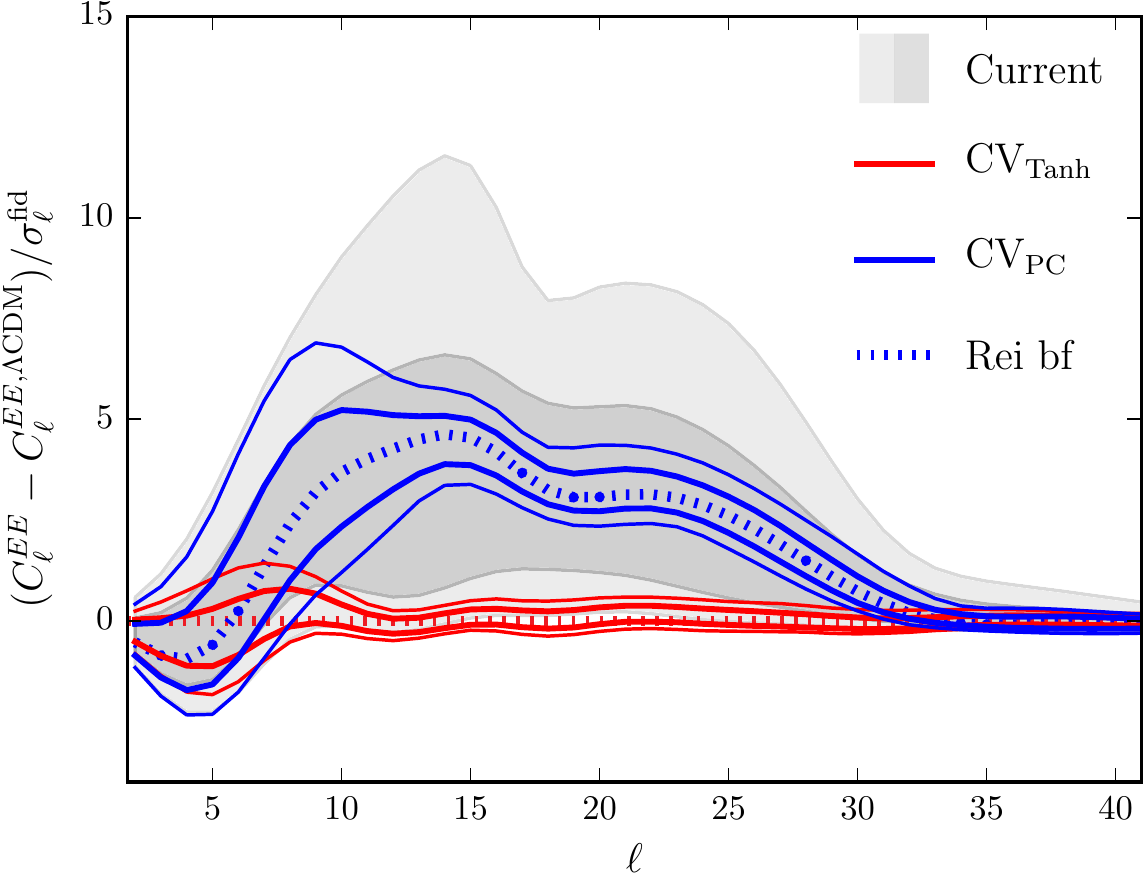}
\caption{\footnotesize $EE$ power spectrum constraints for
additional future CV-limited $EE$
identical to the best fit Tanh model to current data (red) or PC extended model (blue) compared to current Rei constraints (gray) (68\% and 95\% CL)
The {\it Planck} 2015 Rei best fit (blue dotted) is in good agreement with the CV$_{\rm PC}$ means. Note that the Tanh bf model is the horizontal line at zero.}
\label{fig:futureReiClEE}
\end{figure}

\subsection{Models and CV-limited Data}

A comparison of $C_\ell^{EE}$ posteriors from $\Lambda$CDM versus the Rei model (shown in the top panel of Fig.~\ref{fig:ClEElin})  indicates that additional data in the multipole range $\ell = 14-30$  may substantially improve
reionization constraints and more definitively test the steplike ``Tanh" reionization history.  In order to quantitatively study the effects of new data, we post-process our Rei chains assuming that future measurements return values of $C_\ell^{EE}$ that are identical to either the {\it Planck} 2015 $\Lambda$CDM (Tanh) or Rei best fit. Importance sampling is performed by reweighting all samples by a factor of the new likelihood $\mathcal{L}_{\rm future}$ given by:
\begin{align*}
  -2\ln{\mathcal{L_{\rm future}}} = \sum_\ell (2\ell + 1)
  \left(\frac{\hat{C}^{EE}_\ell}{C^{EE}_\ell} -\ln\frac{\hat{C}^{EE}_\ell}{C^{EE}_\ell} -1\right),
\end{align*}
\noindent where $\hat{C}_\ell^{EE}$ is the future measurement, $C_\ell^{EE}$ is the model and the sum runs over values of $\ell$ where we assume new data is available. For the $\ell-$range of new data we take $\ell = 14-30$ which gives insight into how much we can eventually learn about reionization from a CV-limited $EE$ spectrum.
Different assumptions on future data and the models are summarized in Table~\ref{tab:futuremodels}.
The current data itself in this range provide negligible constraints in comparison for any of these choices
 and so this procedure does not
significantly double count information.

\setlength{\tabcolsep}{6pt}
\begin{table}
\begin{tabular}{|c|r|r|r|}
  \hline
&   \multicolumn{1}{c|}{Current} &  \multicolumn{1}{c|}{CV$_{\rm Tanh}$} &  \multicolumn{1}{c|}{CV$_{\rm PC}$} \\
 \hline \hline
$t_1$ & $ 0.002\pm  0.053$  & $-0.090\pm  0.013$  &  $0.027\pm 0.013$ \\
$t_2$ & $-0.029\pm  0.101$  & $-0.081\pm  0.040$  &  $0.007\pm 0.063$ \\
$t_3$ & $ 0.018\pm  0.127$  & $ 0.114\pm  0.064$  &  $0.015\pm 0.076$ \\
$t_4$ & $-0.012\pm  0.143$  & $-0.114\pm  0.098$  &  $0.045\pm 0.110$ \\
$t_5$ & $ 0.026\pm  0.142$  & $ 0.070\pm  0.133$  &  $0.023\pm 0.132$ \\
  \hline
\end{tabular}
  \caption{\footnotesize Mean and standard deviation of the amplitudes of the reionization PCs.
  Note that the CV$_{\rm PC}$ case has larger errors compared to CV$_{\rm Tanh}$ due to the larger values of ClEE as simulated future data.}
   \label{tab:futureerrors}
\end{table}

\begin{figure}
\centering
\includegraphics[width=0.45\textwidth]{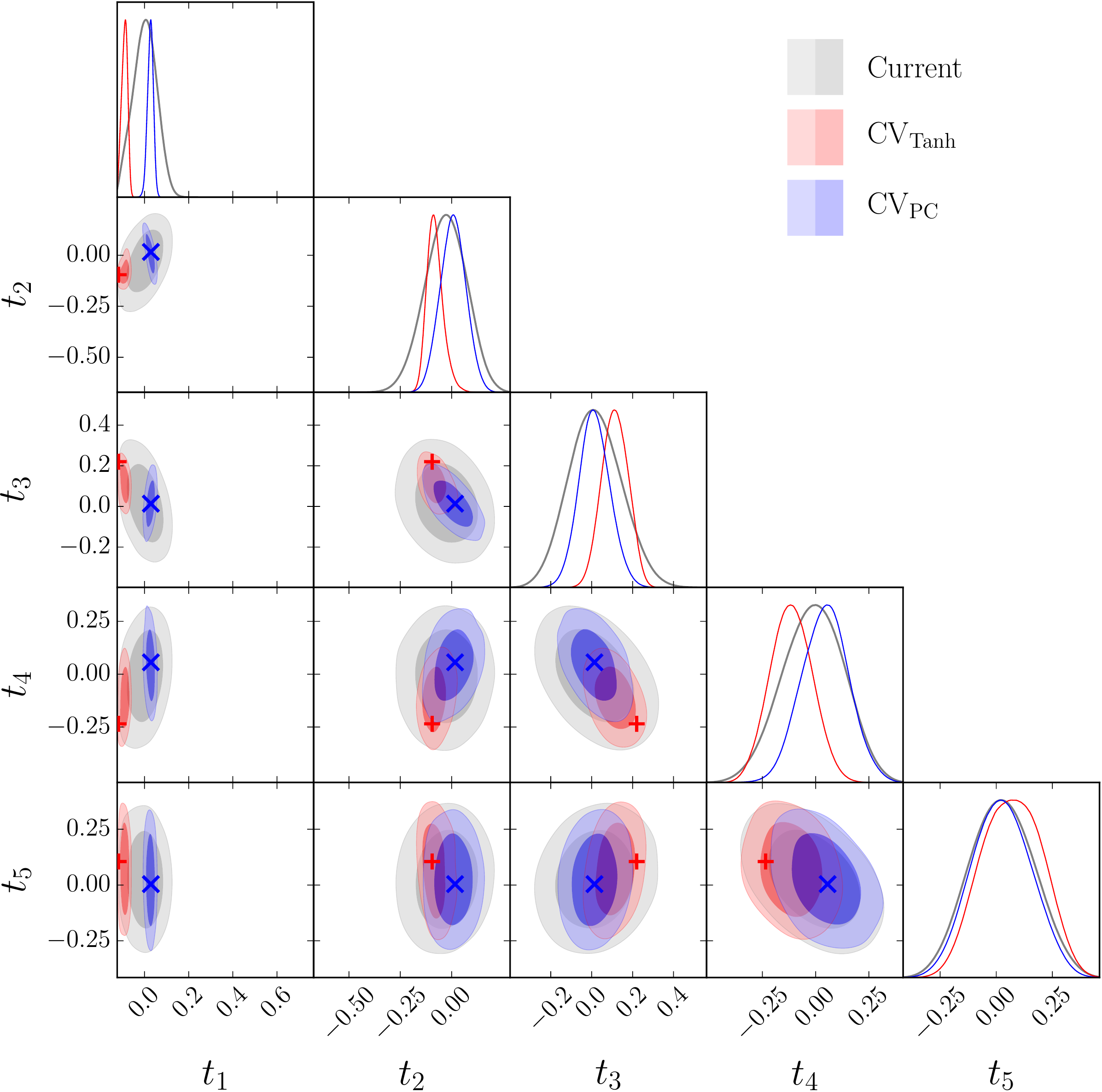}
\caption{\footnotesize Reionization PC constraints with current data combined with future cosmic variance limited $EE$ data in the range $\ell=14-30$ as in Fig.~\ref{fig:futureReiClEE}.
The red $\times$ and blue $+$ indicate these assumed models.  In the $\Lambda$CDM case, the combined data posteriors are not centered on the assumed model
for the future data  since the  current data alone (gray) disfavor the Tanh reionization history.
}
\label{fig:future2DReiall}
\end{figure}
\subsection{Improved Parameter Estimates}

Assuming additional CV limited data severely limits the allowed $C_\ell^{EE}$ model space compared with
current constraints. This is illustrated in Fig.~\ref{fig:futureReiClEE} which shows that the two models are clearly $\rightarrow$ easily distinguishable with a significance  greater than $95\%$ CL over the whole multipole range $10\leq\ell\leq 30$. As anticipated, the extended reionization history model is very strongly disfavored for future data concordant with the Tanh model and vice versa.

A similar story is visible in $t_a$ space (see Fig.~\ref{fig:future2DReiall}) with $t_1=-0.090\pm 0.013$ for CV$_ {\rm Tanh}$ and $t_1 = 0.027\pm 0.013$ for CV$_{\rm PC}$ so that the means are separated by $\sim 9\sigma$. This figure also illustrates the bias of current data against a steplike reionization history. Note that the crosses `$\times$' and plusses `$+$' show the true $t_a$ values for the future data in each case.
The constraints in the $\Lambda$CDM case are consistently displaced toward the Rei best fit points whereas the converse constraints are not.
A summary of the means and standard deviations of the reionization parameters is shown in Table~\ref{tab:futureerrors}.

Assuming a CV-limited measurement at a single $EE$ multipole at for example $\ell = 14$ still yields a separation in $t_1$ sufficient to distinguish between a steplike and extended reionization models. For the CV$_ {\rm Tanh}$ case, we have $t_1= -0.072 \pm 0.022$ while for the CV$_{\rm PC}$ case, we have $t_1 = 0.031\pm 0.023$ such that their means are separated by $\sim 4.5\sigma$.

\begin{figure}
\centering
\includegraphics[width=0.45\textwidth]{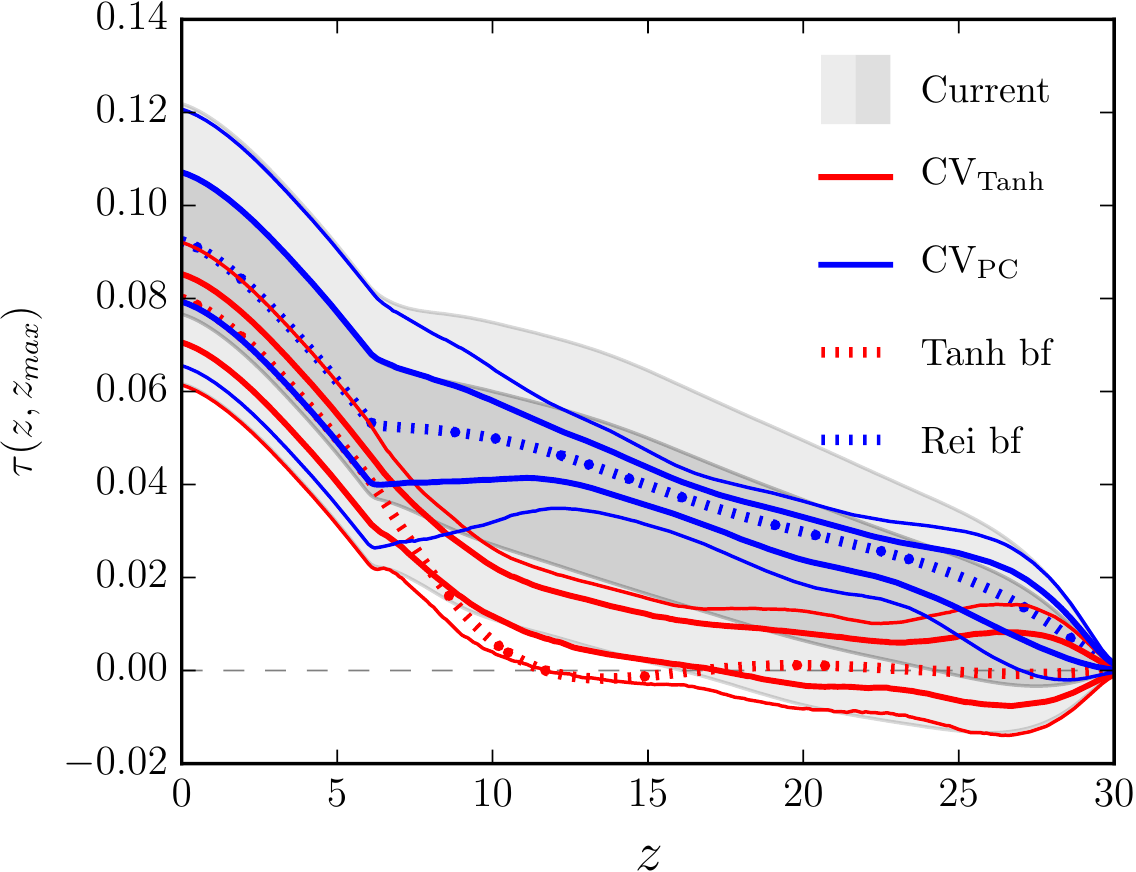}
\caption{\footnotesize Cumulative optical depth $\tau(z, z_{\rm max})$ constraints with future CV-limited $EE$ data as in Fig.~\ref{fig:futureReiClEE}. For the $\Lambda$CDM case, the current data slightly biases the joint constraints toward high-$z$ optical depth ($z\gtrsim 11$) compared with the steplike ionization history assumed for the  additional data. }
\label{fig:futureReiTauZ}
\end{figure}

\begin{figure}[t]
\centering
\includegraphics[width=0.45\textwidth]{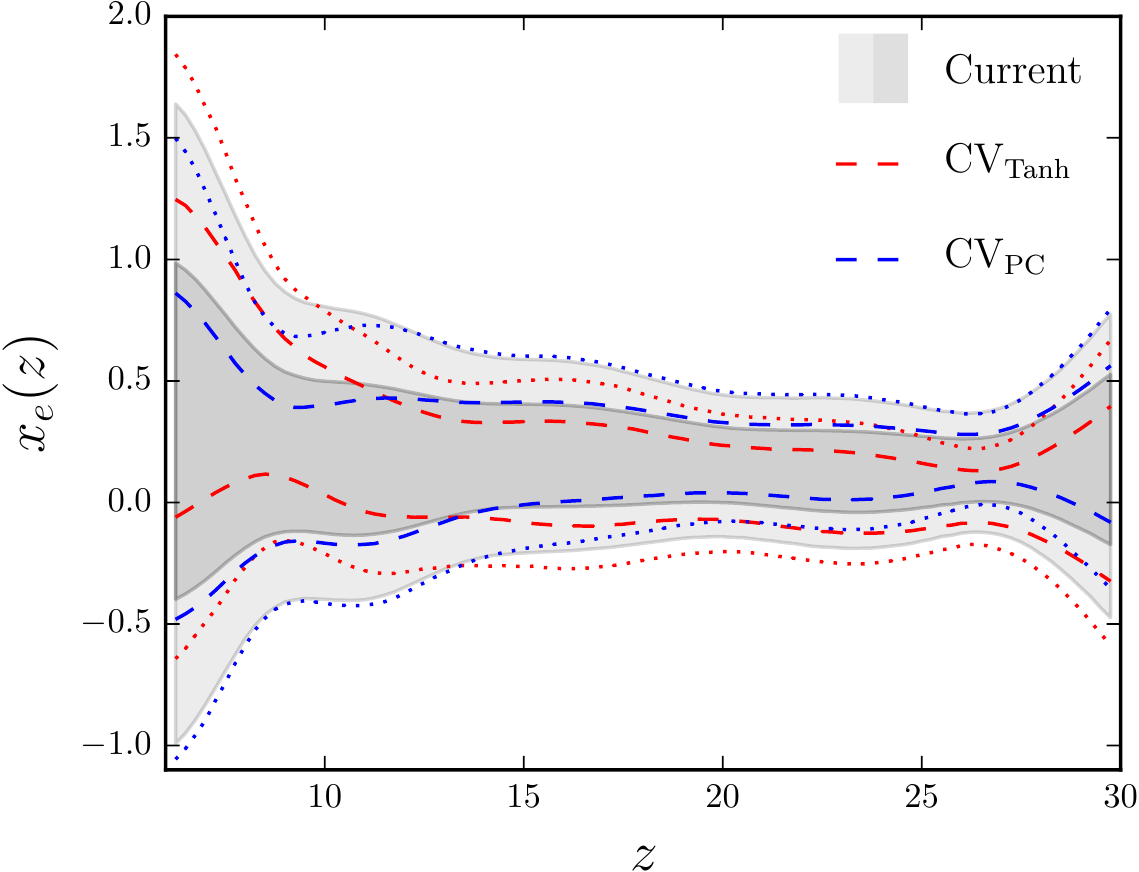}
\caption{Free electron fraction $x_e(z)$ constraints including future CV-limited $EE$ data
 as in Fig.~\ref{fig:futureReiClEE}. Note that even simulated CV-limited $EE$ data is not sufficient to constrain the ionization fraction despite improved cumulative optical depth constraints shown in Fig.~\ref{fig:futureReiTauZ}.}
\label{fig:futureXe}
\end{figure}

\subsection{Optical Depth from High Redshift}
Improved constraints on $t_a$ also imply improved constraints on $\tau(z, z_{\rm max})$. These are shown in Fig.~\ref{fig:futureReiTauZ}.
The two cases are distinguished at high statistical significance through the $z=10-25$ range.
Notably, assuming future data is CV$_ {\rm Tanh}$ the constraints still allow for some contributions to $\tau(z,z_{\rm max})$ from $z\gtrsim 11$ as opposed to its assumed Tanh model.
This reflects the fact that  current data inherently prefers a non-zero contribution at these redshifts.

\subsection{Reionization Fraction Constraints}

Finally, for completeness we consider constraints on the free electron fraction $x_e(z)$ as derived from the 5 $t_a$ constraints as shown in Fig.~\ref{fig:futureXe}.   As mentioned in the main text, this visualization of the constraints emphasizes the most poorly constrained parameters
which allow high frequency oscillations in $x_e$.  Even with CV-limited data, the improvement in $t_4$ and $t_5$ is not
enough to make the reconstructed $x_e$ a good representation of the constraints.  In particular $t_5$ is mainly constrained by the
Doppler contributions to the temperature power spectrum rather than polarization.    This representation would give the misleading impression that
the Tanh and PC models are indistinguishable despite  the clear separation of the models in $t_1$.

Moreover, had we included more than 5 $t_a$ parameters, this representation would become even more misleading.
Since the $x_e(z)$ reconstruction highlights the least constrained  direction in $t_a$ space, the contours would eventually reflect only the priors
and not the data.   This means that the CMB is not capable of constraining the ionization fraction at any one particular redshift despite good constraints on the cumulative $\tau(z,z_{\rm max})$ for all $z$ moderately less than $z_{\rm max}$.

\begin{acknowledgements}
GO and CD were supported by the Dean's Competitive Fund for Promising Scholarship at Harvard University. CH was supported by Jet Propulsion Laboratory, California Institute of Technology, under a contract with the National Aeronautics and Space Administration.
WH was supported by  U.S. Dept. of Energy contract DE-FG02-13ER41958, NASA ATP NNX15AK22G and the Simons Foundation.
 Computing resources were provided by the University of Chicago Research Computing Center through the Kavli Institute for Cosmological Physics at the University of Chicago.
\end{acknowledgements}

\vfill
\clearpage

\bibliography{ReiovsInf}

\end{document}